\documentclass[aps,prl,reprint,groupedaddress,showpacs,superscriptaddress,floatfix]{revtex4-1}
\usepackage{graphicx,amsmath,amsthm,amssymb,color}

\setcitestyle{super}
\definecolor{darkblue}{rgb}{0.184, 0.192, 0.584}
\usepackage[colorlinks=true,linkcolor=black,
urlcolor=black,anchorcolor=darkblue,
citecolor=blue]{hyperref}

\begin{document}

	\title{Experimental realization of phase-controlled dynamics with hybrid digital-analog approach}
	
	\author{Ziyu Tao}
	\thanks{Z. T. and L. Z. contributed equally to this work.}
	\author{Libo Zhang}
	\thanks{Z. T. and L. Z. contributed equally to this work.}
	\author{Xiaole Li}
	\author{Jingjing Niu}
	\author{Kai Luo}
	\author{Kangyuan Yi}
	\author{\\Yuxuan Zhou}
	\author{Hao Jia}
	\author{Song Liu}
	\email{lius3@sustech.edu.cn}
	\author{Tongxing Yan}
	\email{yantx@sustech.edu.cn}
	\author{Yuanzhen Chen}
	\email{chenyz@sustech.edu.cn}
	\author{Dapeng Yu}

	\affiliation{Shenzhen Institute for Quantum Science and Engineering and Department of Physics, Southern University of Science and Technology, Shenzhen 518055, China}
	
	\affiliation{Guangdong Provincial Key Laboratory of Quantum Science and Engineering, Southern University of Science and Technology, Shenzhen, 518055, China}
	
	\affiliation{Shenzhen Key Laboratory of Quantum Science and Engineering, Southern University of Science and Technology, Shenzhen 518055, China}
	
	\date{\today }

	\begin{abstract}
		{Quantum simulation can be implemented in pure digital or analog ways, each with their  pros and cons. By taking advantage of the universality of a digital route and the efficiency of  analog simulation, hybrid digital-analog approaches can  enrich the possibilities for quantum simulation. We use a unique hybrid approach to experimentally perform a quantum simulation of  phase-controlled dynamics resulting from a closed-contour interaction (CCI) within certain multi-level systems in superconducting quantum circuits. Due to symmetry constraints, such systems cannot host an inherent CCI. Nevertheless, by assembling analog modules corresponding to their natural evolutions and specially designed digital modules constructed from standard quantum logic gates, we can bypass such  constraints and realize an effective CCI in these systems. Based on this realization, we  demonstrate a variety of related and interesting phenomena, including phase-controlled chiral dynamics, separation of chiral enantiomers, and a new mechanism to generate entangled states based on CCI.}
	\end{abstract}

	\vskip 0.5cm
	\maketitle
	
	
	Digital quantum simulation relies on decomposition of the evolution of a targeted Hamiltonian into a sequence of discrete quantum logic gates. \cite{Georgescu2014,Barends2015,Barends2016} While in principle this can be done for an arbitrary quantum system,\cite{Lloyd1996} it often requires an intimidating number of gate operations with high precision. Analog approaches  exploiting the continuous nature of quantum evolutions may often be more efficient, \cite{Roushan2016,Wang2019a,Cai2019,Liu2020} but usually must be designed on an $ad$ $hoc$ basis. Hybrid digital-analog quantum simulation has thus been proposed to combine the universality of digital approaches with analog efficiency. \cite{Mezzacapo2015,Lamata2017,Lamata2018,Parra-Rodriguez2020} The flexibility in engineering and assembling digital and analog modules generates abundant possibilities for quantum simulation that are hardly available otherwise. For example, in a simulation of the quantum Rabi model,\cite{Rabi1936} a deep-strong coupling that is inaccessible to pure analog or digital approaches could be realized via a hybrid method. \cite{Langford2017}
	
	In this work, we show that by employing a hybrid method, one can  perform quantum simulations that otherwise cannot be implemented on a given platform. In particular, we demonstrate  phase-controlled quantum dynamics and related phenomena via closed-contour interaction (CCI) in superconducting quantum circuits, which was originally forbidden by certain symmetry-imposed selection rules. The simplest realization of CCI involves a three-level system. Such systems with two of the three possible transitions being coherently driven have been widely researched for both fundamental interest and promising applications in areas such as quantum sensing \cite{Phillips2001,Vanier2005} and quantum information processing.\cite{Cirac1997} By opening the third transition, the three levels form a loop with a CCI, which leads to fundamentally new quantum phenomena, including phase-dependent coherent population trapping,\cite{Kosachiov1992}  phase-controlled dynamics, \cite{Buckle1986} and coherence protection.\cite{Barfuss2018} A closed-loop configuration can also be used in the detection and separation of enantiomers, \cite{Kral2001,Kral2003,Ye2018} i.e., chiral molecules with left ($L$) and right ($R$) handedness, which has long been a challenging problem in chemistry. \cite{Knowles2002}
	
	In practice, the implementation of CCI is often hindered by selection rules for transitions imposed by symmetry constraints in realistic systems. Common practice in overcoming this problem includes the simultaneous use of multiple drivings of different types  (e.g., both electric and magnetic dipole transitions)\cite{Barfuss2018} or  high-order processes such as a two-photon transition. \cite{Vepsalainen2019,Vepsalainen2020} Here, we first show that in a three-level system subject to such selection rules, one can engineer the system Hamiltonian by assembling two digital and one analog module to induce a CCI with only two coherent drivings of the same type. Phenomena related to CCI, such as  phase-controlled chiral dynamics, are observed. By making such driving fields time-dependent, we are able to demonstrate a proposed scheme to separate chiral molecules with high fidelity, \cite{Vitanov2019} and we can extend our technique to more complex systems. Specifically, we propose and realize a new scheme to generate entangled states using a CCI across two coupled superconducting qubits.
	
\begin{figure}
		\centering
		\includegraphics[width =0.48\textwidth]{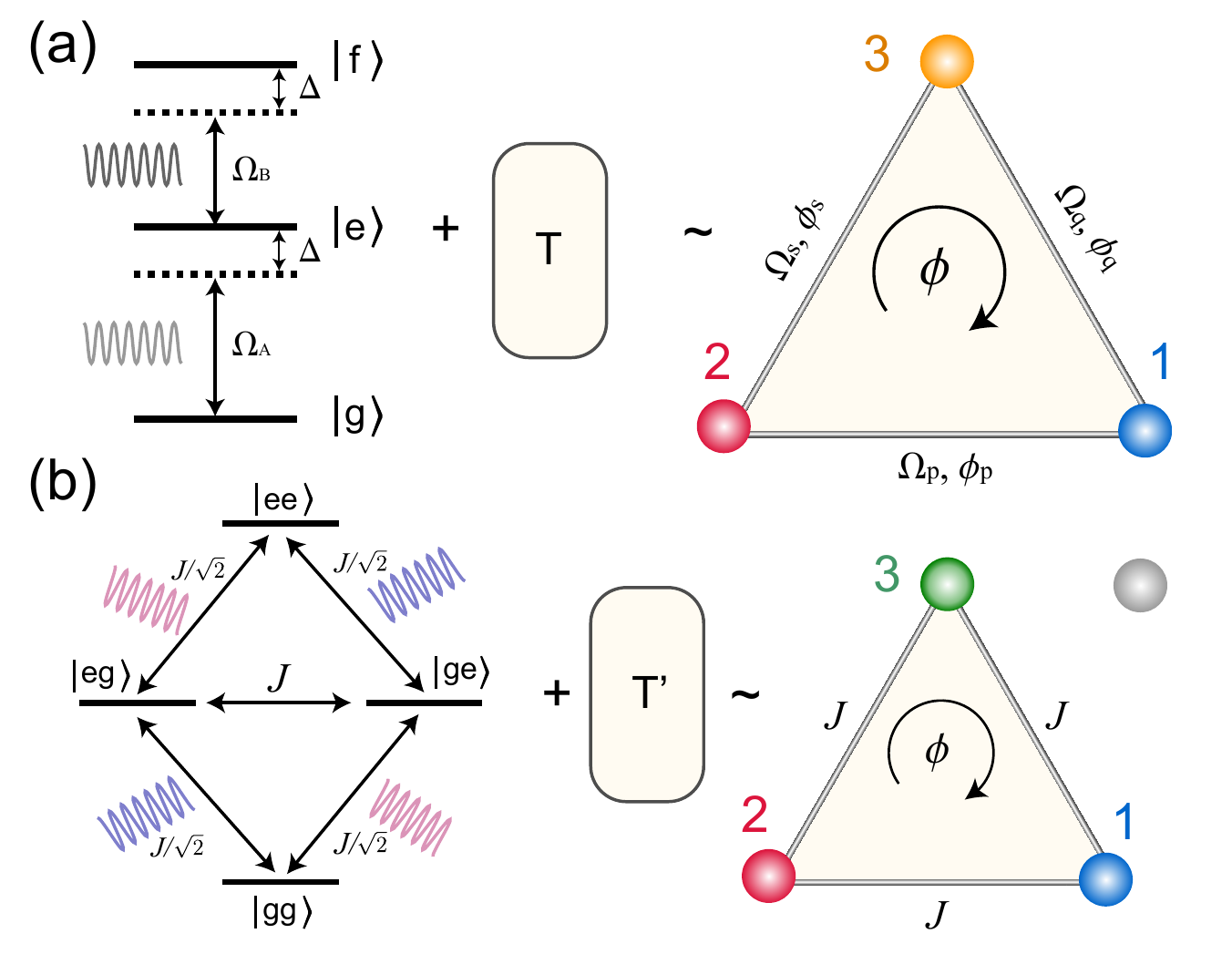}
		\caption{Realization of CCI with a hybrid digital-analog approach. (\textbf{a}) A three-level system (a qutrit) driven by two detuned external fields (described by a Hamiltonian of $H_0$), when combined with specially designed digital modules (the $T$ block) constructed from discrete quantum gates, can be used to realize a new Hamiltonian $H$ hosting an inherent CCI: $T e^{-iH_0 t} T^{\dagger} \equiv e^{-iHt}$, with a gauge-invariant phase $\phi$. For consistency with the literature, we relabel the states of the qutrit as ${\vert g\rangle, \vert e\rangle, \vert f\rangle} = {1, 2, 3}$. (\textbf{b}) In a similar way, combining the natural evolution of two resonant qutrits driven by two external fields (with identical amplitudes and phases, indicated by different colors) with certain digital modules can result in a CCI in a subspace of the system. Here 1, 2, and 3 correspond to $\vert gg\rangle$, $\vert ge\rangle$, and $\vert eg\rangle$, respectively. The gray sphere beside the state of 3 represents a dark state that is decoupled from the evolution of the system (see Methods).}\label{Fig1}
	\end{figure}

	\vspace{10pt}
	\noindent
	\textbf{Results}\\
	\textbf{Realization of CCI.}
	Consider a three-level system composed of three states $\{\vert g\rangle, \vert e\rangle, \vert f\rangle\}$. The system is coherently driven by two external fields of the same type, such as electric-dipole allowed transitions, that correspond to $\vert g\rangle \leftrightarrow \vert e\rangle$ and $\vert e\rangle \leftrightarrow \vert f\rangle$. The effective Hamiltonian of the system under rotating-wave approximation is given by
	\begin{equation}\label{H_0}
	H_0 = \frac{\hbar}{2}\begin{pmatrix}
	-\Delta_{A}&\Omega_{A}^*&0\\
	\Omega_{A}&0&\Omega_{B}^*\\
	0&\Omega_{B}& \Delta_{B}\\
	\end{pmatrix},
	\end{equation}
	where $\Omega_{A,B}$ and $\Delta_{A,B}$ are the amplitudes and detunings, respectively, of the two external driving fields (see Fig. \ref{Fig1}(a)).
	
	\begin{figure}
		\centering
		\includegraphics[width =0.48\textwidth]{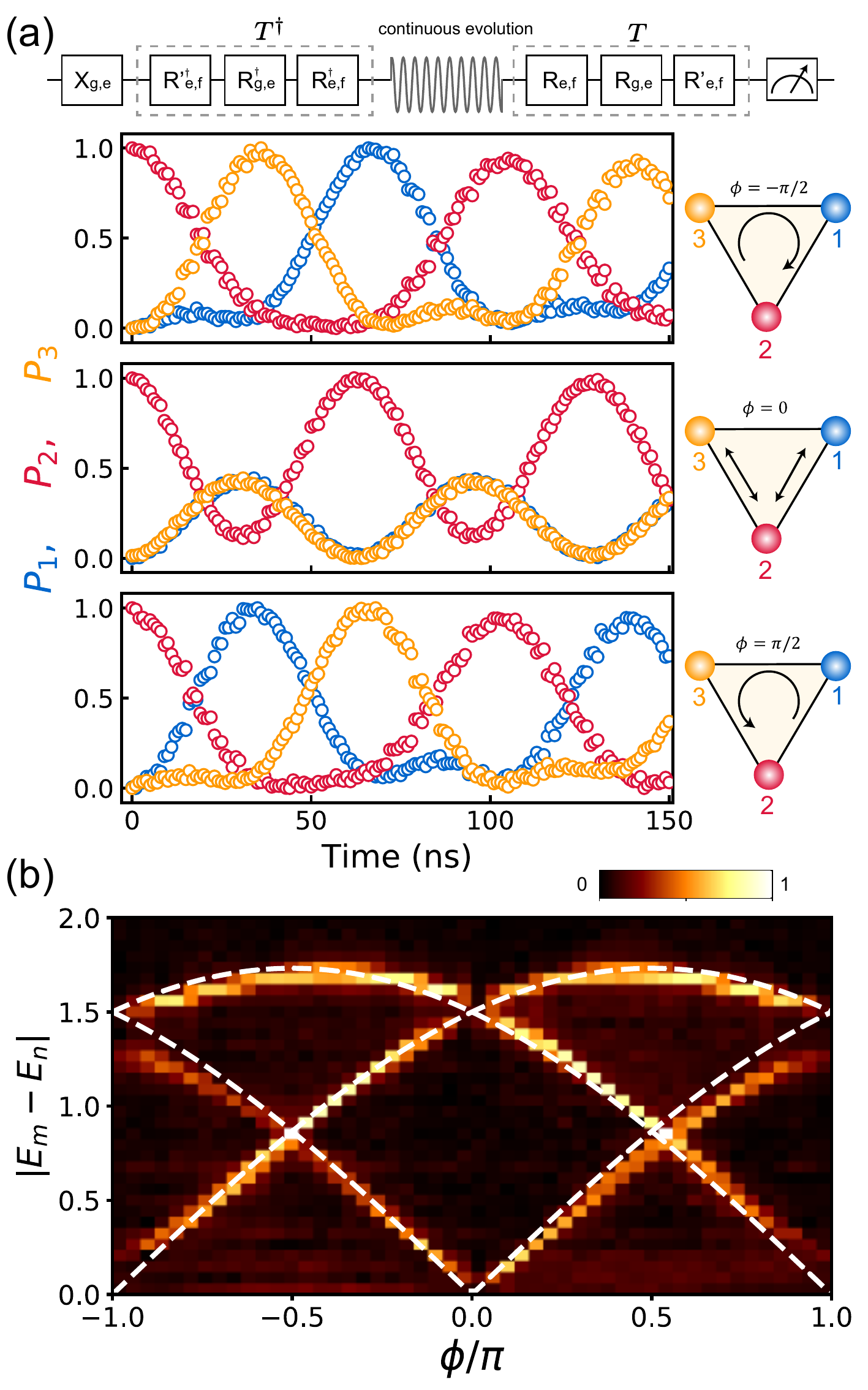}
		\caption{Phase-controlled quantum dynamics resulting from CCI in a single qutrit (see Fig. \ref{Fig1}(a)). (\textbf{a}) Upper part: flowchart of the experiment, including a block for initialization ($X_{g,e}$), a digital module of $T^{\dagger}$ composed of three gate operations, an analog module of the natural evolution driven by $H_0$, and another digital module $T$, followed by projection measurements that yield the three populations of $P_{1,2,3}$. Lower part: $P_{1,2,3}$ as functions of the timespan of the intermediate natural evolution for three  values of the gauge-invariant parameter $\phi$. (\textbf{b}) Energy spectrum of the Hamiltonian of $H$ in Eq. (\ref{eq_hamiChiral3}), obtained via discrete Fourier transform of the measured populations. It is shown in the form of $\vert E_m - E_n \vert$, where $E_k$ are the eigenenergies of $H$, and $m,n\in \{1,2,3\}$. Dashed white lines represent  theoretical predictions. }\label{Fig2}
	\end{figure}	
	
	If the system assumes a restrictive symmetry, then the third transition $\vert g\rangle \leftrightarrow \vert f\rangle$ of the same type is forbidden. Even in systems of less restrictive symmetry (e.g., artificial atoms such as superconducting qubits), the amplitude of such transitions is usually vanishingly small. \cite{Koch2007} Previously, a third driving of a different type or of the same type but of higher order was used to close the loop to form a CCI.  \cite{Vepsalainen2019,Vepsalainen2020} We take a different approach. By combining an analog module corresponding to the evolution driven by $H_0$ with two digital modules that are unitary operators constructed from standard quantum gates, we effectively transform the original Hamiltonian $H_0$ to the following form (see  Methods for details):
	\begin{equation}\label{eq_hamiChiral3}
	H = \frac{\hbar}{2}
	\begin{pmatrix}
	0 & \Omega_p e^{-i\phi_p} & \Omega_q e^{i\phi_q} \\
	\Omega_p e^{i\phi_p} & 0 & \Omega_s e^{-i\phi_s} \\
	\Omega_q e^{-i\phi_q} & \Omega_s e^{i\phi_s} & 0
	\end{pmatrix}.
	\end{equation}
	To arrive at the above form, we set the amplitudes and detunings of the two drivings to  $\Omega_A = \left[ \Omega_p e^{i\phi_p}  + \Omega_s e^{i(\phi_q-\phi_s)} \right] /\sqrt{2}$, $\Omega_B = \left[ -\Omega_p e^{i(\phi_q-\phi_p)} + \Omega_s e^{i\phi_s} \right]/\sqrt{2}$, and $\Delta_{A,B}=-\Omega_q$.

	This new Hamiltonian differs from $H_0$ in that it naturally contains nonzero amplitudes for all three possible transitions, and the magnitudes and phases of all three amplitudes can be adjusted independently (see Fig. \ref{Fig1}(a)). Therefore,  inherent CCI dynamics can be expected for such a Hamiltonian. In the case of equal and constant magnitudes, $\Omega_{p,q,s} \equiv \Omega$, the population dynamics are strongly dependent on the phases $\phi_{p,s,q}$ of the driving fields, through a gauge-invariant global phase $\phi = \phi_p + \phi_s- \phi_q$. We will show an experimental demonstration of such CCI dynamics.

	We  used Xmon-type  superconducting qutrits in our experimental work. In this kind of artificial atom, the transitions of $\vert g\rangle \leftrightarrow \vert e\rangle$ and $\vert e\rangle \leftrightarrow \vert f\rangle$ are electric-dipole allowed, whereas the transition $\vert g\rangle \leftrightarrow \vert f\rangle$ of the same type has a vanishingly small amplitude. \cite{Koch2007} Two external microwave driving fields in the forms described above ($\Omega_{A,B}$) are applied to the qutrit, with $\Omega_{p,q,s} \equiv \Omega$ and three independently adjustable phases $\phi_{p,q,s}$. Details of the experimental setup can be found in the Supplemental Materials.

	\vspace{10pt}
	\noindent
	\textbf{CCI dynamics.}
	We first study the CCI dynamics of the system by measuring its time evolution at different values of $\phi$. Figure \ref{Fig2}(a) shows the temporal sequence of operations. The system is initialized in the first excited state of $\vert \psi(t=0)\rangle = \vert e\rangle$ by a standard $X$ gate. A digital module containing three quantum gates is  applied to the qutrit, followed by an analog evolution driven by $H_0$ with two control parameters: the time span and the gauge-invariant phase $\phi$. Another digital module, which is the Hermitian conjugate of the first digital module, is applied, followed by projection measurements that yield populations of all three states. As discussed previously, the combined effect of the middle three blocks is to subject the system to evolve under a new Hamiltonian $H$ as in Eq. (\ref{eq_hamiChiral3}): $e^{-i H t/\hbar } \equiv T e^{-i H_0 t/\hbar } T^{\dagger}$.

	The gauge-invariant phase $\phi$ assumes a role as the flux of a synthetic magnetic field, which controls the dynamics of the system. At $\phi=0$, the populations evolve in time with a symmetric pattern without a preferred direction of circulation (middle panel,  Fig. ~\ref{Fig2}(a)). Such symmetry in the circulation pattern is not observed for values of $\phi$ that are not integers of $\pi$. Two examples corresponding to $\phi = \pm \pi/2$ are shown in Fig. \ref{Fig2}(a). In each case, a circulation of certain chirality is observed: clockwise for $\phi = -\pi/2$ and counterclockwise for $\phi = \pi/2$. Such differences are rooted in the symmetry of the system upon time reversal. An examination of the time-reversal symmetry (TRS) in a strict sense requires reversing the flow of time, which is of course not experimentally feasible. However, the periodicity presented in the evolutions shown in Fig. \ref{Fig2}(a) allows for a practical definition of the TRS: $\psi(t) = \psi(T_0-t)$, where $T_0$ is the period of a given evolution \cite{Roushan2016}. By comparing the evolutions from $t = 0$ forward and from $t = T_0$ backward, Fig.~\ref{Fig2}(a) shows that the TRS is preserved for $\phi = 0$, but broken for $\phi = \pm\pi/2$.
	
	In addition to demonstrating the phase-controlled dynamics under CCI, we  mapped out the electronic structure of the system as a function of $\phi$. The eigenenergies of $H$ are given by $E_{k} = \Omega \cos[\phi/3 - \varphi_0 (k+1) ]$, with $k\in \{1,2,3\}$ and $\varphi_0 = 2\pi/3$. A Fourier transformation of the measured populations can reveal the energy differences $\vert E_m - E_n \vert$ with $m,n\in \{1,2,3\}$ and $m\ne n$, as shown in Fig.~\ref{Fig2}(b), which agree with the simulated results using $H$ in Eq. (\ref{eq_hamiChiral3}). The anti-crossings at $\phi = \pm \pi$ in the spectrum can be explained by the slight detuning of the coherent drives and environmental fluctuations. \cite{Barfuss2018}
	
	\begin{figure}[!t]
		\centering
		\includegraphics[width =0.48\textwidth]{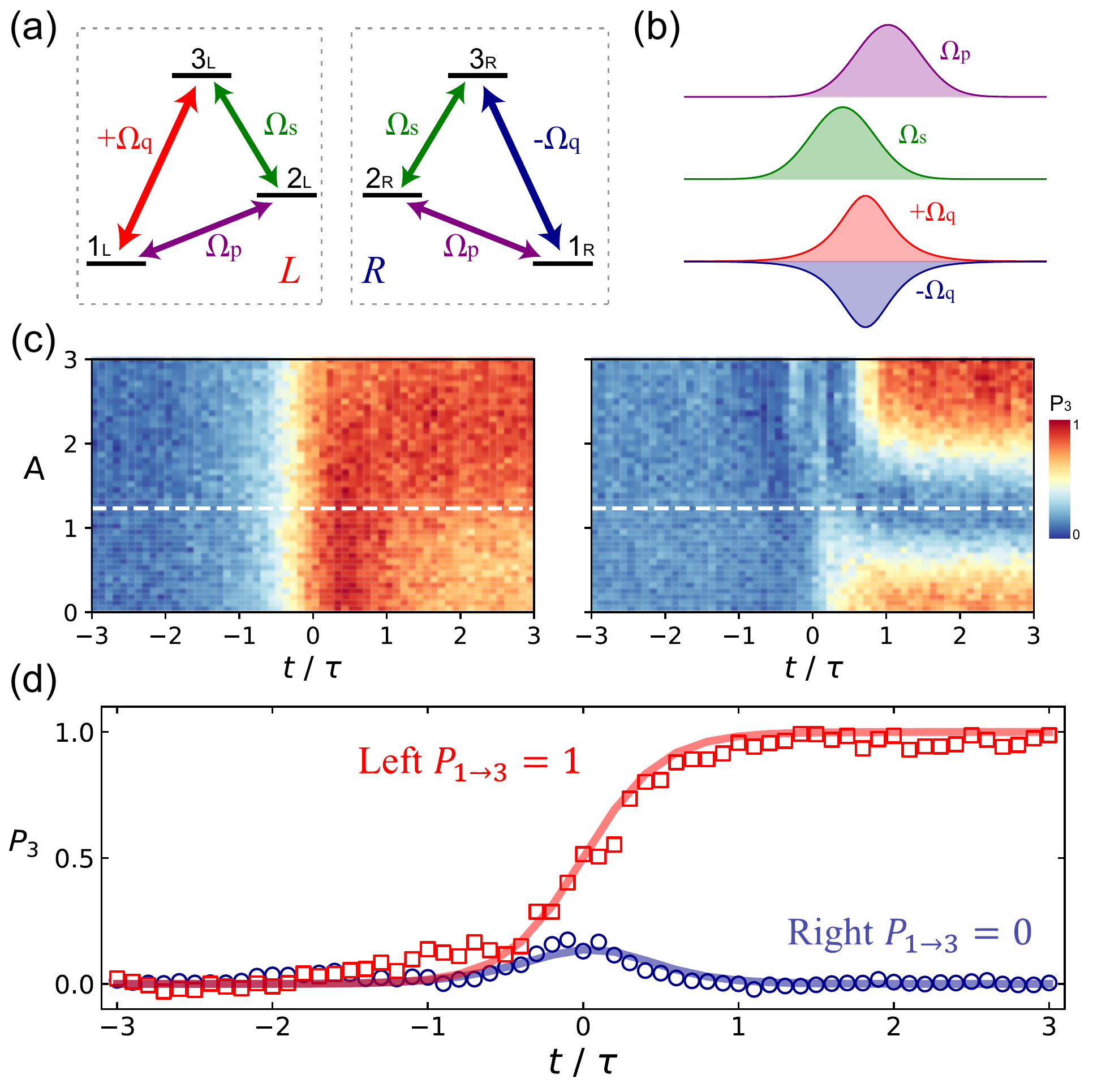}
		\caption{Chiral separation via CCI. (\textbf{a}) Coupling schemes of chiral molecules with $L$ and $R$ handedness. Identical drivings result in a difference of $\pi$ in the overall phase of the loop, indicated here as different couplings ($\pm\Omega_q$) between the states of 1 and 3. (\textbf{b}) Pulse sequence for the driving fields $\Omega_p(t)$, $\Omega_s(t)$, and $\pm\Omega_q(t)$. (\textbf{c}) Measured population $P_3$ versus time and pulse area $A\pi$ for $L$ (left panel) and $R$  (right panel) handedness, where the initial state $\vert \psi_0\rangle = \vert 1\rangle$. The maximum population contrast is obtained when $A\approx 1.23$ (indicated by the white dashed lines). $t=0$ corresponds to the moment when the $\pm\Omega_q$ pulse reaches its maximum magnitude. (\textbf{d}) The population $P_3$ as a function of time for $A=1.23$, showing that the transfer to the state of 3 is nearly perfect for $L$ handedness, but completely suppressed for $R$ handedness.}\label{Fig3}
	\end{figure}
	
	\begin{figure*}
		\centering
		\includegraphics[width =1\textwidth]{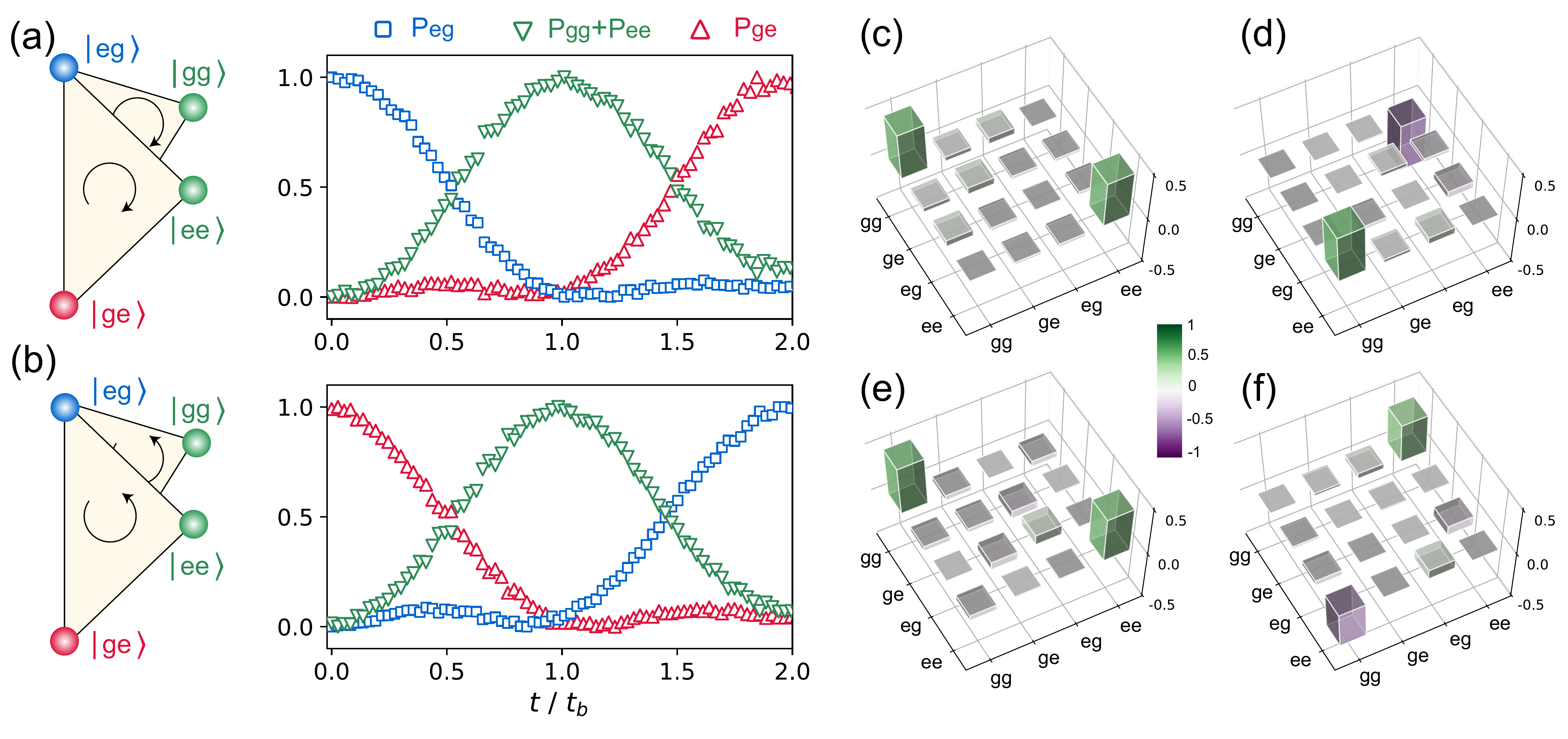}
		\caption{Generation of entangled states using CCI. (\textbf{a-b}) The measured populations of $P_{eg},P_{ge}$, and $P_{gg}+P_{ee}$. The non-entangled states $\vert eg \rangle $ and $\vert ge \rangle $ in (\textbf{a}) and (\textbf{b}) evolve into the entangled states of $\left(\vert gg\rangle \pm i \vert ee\rangle \right)/\sqrt{2}$ within  time  $t_b$ under the maximum TRS breaking condition $\phi = \pm \pi/2$. (\textbf{c}) Real and (\textbf{d}) imaginary parts of the density matrix for the entangled state of $\left(\vert gg\rangle + i \vert ee\rangle \right)/\sqrt{2}$, constructed from data obtained by quantum state tomography. (\textbf{e}) Real and (\textbf{f}) imaginary parts for the state of $\left(\vert gg\rangle - i \vert ee\rangle \right)/\sqrt{2}$.}\label{Fig4}
	\end{figure*}		
	
	\vspace{10pt}
	\noindent
	\textbf{Chiral separation.}
	Beyond constant driving fields, we further consider a closed loop driven by three time-dependent fields $\Omega_p(t)$, $\Omega_s(t)$, and $\Omega_q(t)$, which was proposed to detect and separate enantiomers with $L$ and $R$ handedness by using the phase-sensitive interferometric nature of the closed-loop configuration.\cite{Vitanov2019}

	For a three-level system subjected to a pumping drive $\Omega_p(t)$ ($\vert 1\rangle \leftrightarrow \vert 2\rangle$) and  Stokes drive $\Omega_s(t)$ ($\vert 2\rangle \leftrightarrow \vert 3\rangle$) (see Fig. \ref{Fig3}(a); for consistency with the literature, here we label the three states as $\vert 1\rangle$, $\vert 2\rangle$, and $\vert 3\rangle$), the three eigenenergies and corresponding eigenstates are $\lambda_{\pm} = \pm \sqrt{\Omega_p^2 + \Omega_s^2}$, $\lambda_0 = 0$, and $\vert \chi_\pm \rangle = (\sin \theta \vert 0 \rangle \pm \vert 2 \rangle + \cos \theta \vert 3 \rangle )$, $\vert \chi_0 \rangle =\cos\theta \vert 1 \rangle - \sin \theta \vert 3\rangle$, with $\tan\theta(t)=\Omega_p(t)/\Omega_s(t)$. In the celebrated technique of stimulated Raman adiabatic passage, \cite{Vitanov2017} the two pulses are arranged in a counterintuitive order with the Stokes pulse coming first, and the eigenstate $\vert \chi_0 \rangle$ evolves adiabatically from $\vert 1\rangle$ to -$\vert 3\rangle$ as $\theta$ varies from 0 to $\pi$/2, thus accomplishing a nearly perfect state transfer coherently.

	It has been shown that by adding a counterdiabatic driving $\Omega_q(t)$ ($\vert 1\rangle \leftrightarrow \vert 3\rangle$) to close the loop, the resultant dynamics of the population become dependent on the handedness of the system. \cite{Chen2010,Guery-Odelin2019} In particular, with the same driving fields, the Hamiltonian of the system is $H^{L,R} = \left(\Omega_p \vert 2 \rangle \langle 1\vert + \Omega_s \vert 3 \rangle \langle 2\vert \pm \Omega_q e^{i\phi} \vert 3 \rangle \langle 1\vert \right)+ H.c. $ (Fig.~\ref{Fig3}(a)), where the $+(-)$ sign is for $L(R)$ handedness, and $H.c$ is the Hermitian conjugate. Such a sign difference will result in the  same counterdiabatic driving  doubling or canceling the nonadiabatic coupling presented in the system, depending on its handedness. If $\phi$ is set to  $-\pi/2$, then the populations of the final state, $P_{3}$, of the enantiomers with $L$ and $R$ handedness are different. For example, with carefully chosen values of the pulse areas, the handedness can be efficiently determined by measuring $P_{3}$ alone, where $P_{3} =1 (P_{3} =0 ) $ for $L(R)$ handedness. \cite{Vitanov2019} We note that such a counterdiabatic driving was originally proposed to accelerate various adiabatic processes, but here its major effect is to differentiate the $L$ and $R$ handedness.

	We use pump and Stokes pulses of a Gaussian form in our experiment: $\Omega_p(t) = \Omega_0 e^{-(t-\tau/2)^2/\tau^2}$, $\Omega_s(t) = \Omega_0 e^{-(t+\tau/2)^2/\tau^2}$. Both pulses have a width of $\tau$ and are delayed by the same amount. A third pulse in the form of $\Omega_q(t) = \pm2\dot{\theta}(t)$ is applied, where the $+(-)$ sign corresponds to  $L$($R$) handedness. We prepare the system in an initial state of $\vert \chi_0 \rangle$. As discussed above, for  $L$ handedness, the nonadiabatic transition is canceled by $\Omega_q(t)$ and the system remains in the state $\vert \chi_0 \rangle$, inducing a perfect population transfer from $\vert 1 \rangle$ to $\vert 3\rangle$ with $P_{1\to 3} =1$ as $\theta(t)$ evolves from 0 to $\pi/2$. Conversely, for  $R$ handedness, the nonadiabatic transition doubles, which enables $\vert \chi_0 \rangle \to \vert \chi_\pm \rangle$ and $P_{1\to 3} <1$. Figure~\ref{Fig3}(c) shows the time evolution of $P_3$ with different pulse areas $A\pi$, which is defined as $\int\Omega_{p,s} dt = \Omega_0 \tau \sqrt{\pi}\equiv A\pi$. The driving fields $\Omega_{p,s,q}$ in Fig.~\ref{Fig3}(b) result in a population transfer $\vert 1 \rangle \to \vert 3 \rangle$ for $L$ handedness with $P_{1\to 3} =0.986$, and a suppression of the same transfer for  $R$ handedness  with $P_{1\to 3} =0.003$ when $A\approx 1.23$ (Fig.~\ref{Fig3}(d)).

	\vspace{10pt}
	\noindent
	\textbf{Entanglement generation with CCI.}
	Next, we extend the generation of CCI via pure microwave drivings to a more complex system of two coupled qubits, and further demonstrate a new mechanism of entangling two qubits based on CCI, different from  existing schemes that are widely used in quantum information processing with superconducting quantum circuits.
	
	Consider the four-level system formed by two Xmon superconducting qubits with a coupling strength of $J$ (see Fig. \ref{Fig1}(b)). We apply two transverse resonant driving fields to the two qubits, with an identical amplitude of $J/\sqrt{2}$ and a phase difference of $\phi_a-\phi_b = \phi$. Similar to the single-qubit case discussed above, we combine the natural evolution of such a driven system (an analog module) and a unitary operation $T'$ (two digital modules implemented via standard gate operations) to realize an effective Hamiltonian for a three-state system $\{\vert eg \rangle,\vert ge \rangle,\vert gg \rangle \}$ that can host CCI (see Fig. \ref{Fig1}(b) and Methods). Furthermore, we can generate entangled states of the two qubits  by removing the unitary operation $T'$, since it transforms the entangled state $\vert gg \rangle + e^{i\phi}\vert ee \rangle $ to the ground state $\vert gg \rangle$, and the special form of $T'e^{-i H t}T'^{\dagger}$ used in this work mathematically corresponds to a linear transformation in the Hilbert space.
	
	Specifically, the two-qubit system can be directly transferred from the non-entangled state $\vert eg\rangle$ or $\vert ge\vert$ to the maximum entangled states of $\left(\vert gg\rangle \pm i \vert ee\rangle\right)/\sqrt{2}$ (Fig. \ref{Fig4}(a) and (b)), within a time of $t_b =2\pi/(3\sqrt{3}J) $, under the condition of maximum TRS breaking at $\phi = \pm \pi/2$. The density matrices $\rho_{\pm}$ of the entangled states $\vert\psi_{\pm}\rangle$ characterized by quantum state tomography are given in Fig.~\ref{Fig4}(c)-(f), with fidelities of $F_{+} = 0.963\pm 0.026 $ and $F_{-} = 0.923\pm 0.029 $. The analytical form of the nontrivial two-qubit unitary operator $e^{-iHt_b}$ is given in the Supplementary Information. This new mechanism to generate  entanglement based on chiral CCI dynamics is different from the previous constructions of iSWAP \cite{Schuch2003,Bialczak2010} and controlled-Z gates, \cite{Strauch2003,Yamamoto2010,Ghosh2013} formed by the subspace $\{\vert ge \rangle,\vert eg\rangle\}$ or $\{\vert ee \rangle,\vert fg\rangle\}$ in superconducting qubits.
	
	\vspace{10pt}
	\noindent
	\textbf{Discussion.}
	We have proposed and experimentally demonstrated an effective realization of CCI in genuine three-level systems that do not host CCI inherently due to certain symmetry constraints. By assembling an analog module of the natural evolution governed by their original Hamiltonians with carefully designed digital modules, we can effectively bypass such constraints and establish a CCI without  auxiliary driving signals that are technically challenging to implement. Based on such a CCI, we can demonstrate a variety of interesting related  phenomena such as a phase-controlled chiral dynamics, chiral separation, and a new mechanism to generate entangled states.
	
	The hybrid digital-analog approach used here is essential to our work, since on the one hand the above symmetry constraints  forbid an inherent CCI that would manifest in the analog evolutions of the systems, and on the other hand, a pure digital approach is practically infeasible, as too many quantum gate operations would be required, especially to simulate the natural evolutions of the systems. This work serves as a preliminary demonstration of the enriched possibilities for quantum simulation by the hybrid digital-analog approach. One can reasonably expect, by assembling more sophisticated and ingeniously engineered analog and digital modules, the realm of quantum simulation that is accessible by pure analog or digital approaches can be largely expanded, a welcome development before we realize a universal and fault-tolerant digital quantum computer.

	\vspace{10pt}
	\noindent
	\textbf{Methods}\\
	{\footnotesize
		
		\noindent
		\textbf{Experimental setup.}
		We  used the Xmon-type of superconducting qutrit with a tunable frequency via a bias current on a Z-control line. Microwave pulses are applied to the qutrit via an XY-control line. The state of the qutrit can be deduced by measuring the transmission coefficient $S_{21}$ of the transmission line using a standard dispersive measurement \cite{Wallraff2005}. For the part of experiment involving two qubits, they are coupled via an ancillary qubit that can fine tune the effective coupling strength \cite{Yan2018}. Further details of the samples and measurement setup can be found in the Supplementary Information.
		
		\vspace{10pt}
		\noindent
		\textbf{Effective Hamiltonian of the three-level system.}
		The effective Hamiltonian of the microwave-driven qutrit in a rotating frame described by the operator $U = \vert g \rangle \langle g \vert + \vert e \rangle \langle e \vert e^{i\omega_A t}+ \vert f \rangle \langle f \vert e^{i (\omega_A t+\omega_B t) } $ and under the rotating-wave approximation is given by Eq. \ref{H_0}. The unitary operator $T$ that serves as a digital module is
		\begin{equation}
		T =
		\begin{pmatrix}
		1/\sqrt{2} & 0 & -e^{i \phi_q}/\sqrt{2}\\
		0 & 1 & 0 \\
		e^{-i \phi_q}/\sqrt{2} & 0 & 1/\sqrt{2}
		\end{pmatrix},
		\end{equation}
		which can be constructed from three single-qutrit gates $R_{e,f}(\pi,0)\cdot R_{g,e}(\pi/2,-\phi_q) \cdot R_{e,f}(\pi,\pi)$, where $R_{m,n}(\theta,\phi)$ represents a rotation in the subspace of $\{\vert m \rangle, \vert n\rangle\}$:
		\begin{equation}
		R_{m,n}(\theta,\phi) =
		\begin{pmatrix}
		\cos(\theta/2)& -e^{-i\phi}\sin(\theta/2)\\
		e^{i\phi}\sin(\theta/2)&\cos(\theta/2)
		\end{pmatrix}.
		\end{equation}
		
		The combination of the natural evolution of the original Hamiltonian and the unitary operations gives the effective Hamiltonian $H$ in Eq.~(\ref{eq_hamiChiral3}): $e^{-i H t/\hbar } \equiv T e^{-i H_0 t/\hbar } T^{\dagger} $, which describes a three-level system with CCI.
		
		\vspace{10pt}
		\noindent
		\textbf{Effective Hamiltonian of the four-level system.}
		Consider the four-level system formed by two coupled superconducting qubits with a coupling strength of $J$. We apply two transverse resonant driving fields, one to each qubit, with identical frequency $\omega_a = \omega_b =\omega_{ge}$ and amplitude $ \vert \Omega_{A} \vert= \vert\Omega_{B} \vert = J/\sqrt{2}$, and a phase difference of $\phi_a-\phi_b = \phi$. In a rotating frame described by an operator $U = \left(\vert g \rangle \langle g \vert + \vert e \rangle \langle e \vert e^{i\omega_a t} \right) \otimes  \left(\vert g \rangle \langle g \vert + \vert e \rangle \langle e \vert e^{i\omega_b t} \right) $ and under the rotating-wave approximation, the Hamiltonian is given by
		\begin{equation}
		\begin{split}
		H/\hbar &= J(\cos\phi\sigma^a_x + \sin\phi\sigma^a_y+\sigma^b_x)/\sqrt{2} + J(\sigma^a_x\otimes\sigma^b_x+\sigma^a_y\otimes\sigma^b_y)/2
		\\
		&=
		\frac{J}{\sqrt{2} }
		\begin{pmatrix}
		0&1&e^{-i\phi}&0\\
		1&0&\sqrt{2}&e^{-i\phi}\\
		e^{i\phi}&\sqrt{2}&0&1\\
		0& e^{i\phi}&1&0\\
		\end{pmatrix}
		\end{split}.
		\end{equation}
		Combining the natural evolution governed by this Hamiltonian and a unitary operation defined as
		\begin{equation}
		T' = \frac{1}{\sqrt{2} }
		\left(
		\begin{array}{cccc}
		1& 0 & 0 & e^{-i\phi}\\
		0& \sqrt{2} & 0 & 0 \\
		0& 0 & \sqrt{2} & 0 \\
		-e^{i\phi}& 0 & 0 & 1\\
		\end{array}
		\right)
		\end{equation}
		gives an effective Hamiltonian $H'$ via $e^{-i H' t/\hbar } \equiv T' e^{-i H t/\hbar } T'^{\dagger}$:
		\begin{equation}
		H'= J\left( \vert {eg} \rangle \langle {ge} \vert + \vert {ge} \rangle \langle {gg} \vert + e^{i\phi} \vert {eg} \rangle \langle {gg} \vert + H.c. \right).
		\end{equation}
		This new Hamiltonian describes a three-level system with CCI. If the two unitary operations, $T'$ and $T'^{\dagger}$ are dropped, then Eq. (7) becomes
		\begin{equation}
		\bar{H}'= J\left( \vert \bar{1} \rangle \langle \bar{2} \vert + \vert \bar{2} \rangle \langle \bar{3} \vert + e^{i\phi} \vert \bar{1} \rangle \langle \bar{3} \vert + H.c. \right).
		\end{equation}
		Here, $\{ \vert \bar{1}\rangle ,\vert \bar{2}\rangle ,\vert \bar{3}\rangle \}$ form an invariant triplet subspace of the overall Hilbert space of $\{\vert \bar{1}\rangle ,\vert \bar{2}\rangle ,\vert \bar{3}\rangle,\vert \bar{D}\rangle \} \equiv \left \{\vert eg\rangle ,\vert ge\rangle , \left(\vert gg\rangle +e^{i\phi}\vert ee\rangle \right)/\sqrt{2},\left(\vert gg\rangle -e^{i\phi}\vert ee\rangle \right)/\sqrt{2}\right \} $, and the state of $\vert D\rangle$ is a dark state that is decoupled from the system evolution.}
	
	\bibliography{ref_chiral3}
	\bibliographystyle{apsrev4-1}

	\vspace{10pt}
	\noindent
	\textbf{Acknowledgements}\\
	{This work was supported by the Key-Area Research and Development Program of Guang-Dong Province (Grant No. 2018B030326001), the National Natural Science Foundation of China (U1801661), the Guangdong Innovative and Entrepreneurial Research Team Program (2016ZT06D348), the Guangdong Provincial Key Laboratory (Grant No.2019B121203002), the Natural Science Foundation of Guangdong Province (2017B030308003), and the Science, Technology and Innovation Commission of Shenzhen Municipality (JCYJ20170412152620376, KYTDPT20181011104202253).}

	\vspace{10pt}
	\noindent
	\textbf{Author contributions}\\
	{\small
		Z. T. and L. Z. contributed equally to this work. T. Y. and Z. T. conceived the experiment; Z. T. designed the theoretical protocol and performed the experiment with T. Y. under the supervision of Y. C.; L. Z. designed the superconducting devices used in the experiment, and fabricated them together with Y. Z. and H. J.; T. Y., Z. T., and Y. C. wrote the manuscript together, with inputs from all authors.}

	\clearpage

\pagebreak
\widetext

	\title{SUPPLEMENTARY INFORMATION}

\begin{center}
	\textbf{\large SUPPLEMENTARY INFORMATION}
\end{center}

	\section{Information of superconducting quantum devices and experimental setup}
	
\begin{figure*}[!ht]
		\centering
		\includegraphics[width =0.7\textwidth]{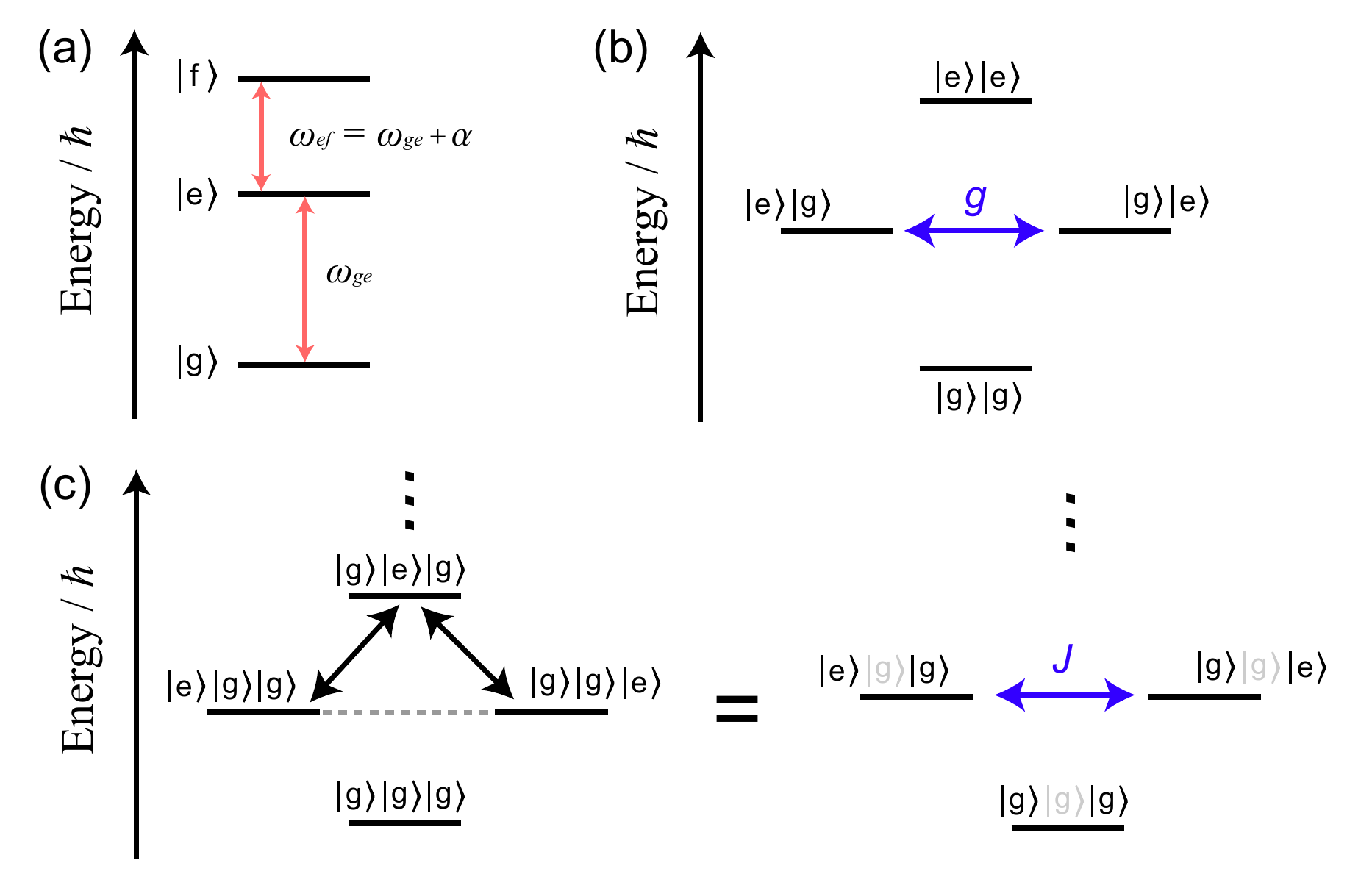}
		\caption{
			(\textbf{a}) Energy levels for a superconducting qutrit,
			(\textbf{b}) two coupled qubits, and
			(\textbf{c}) two qubits effectively coupled through an ancillary qubit, which serves as a tunable coupler.
		}\label{energy}
	\end{figure*}

\begin{figure*}
		\centering
		\includegraphics[width =0.9\textwidth]{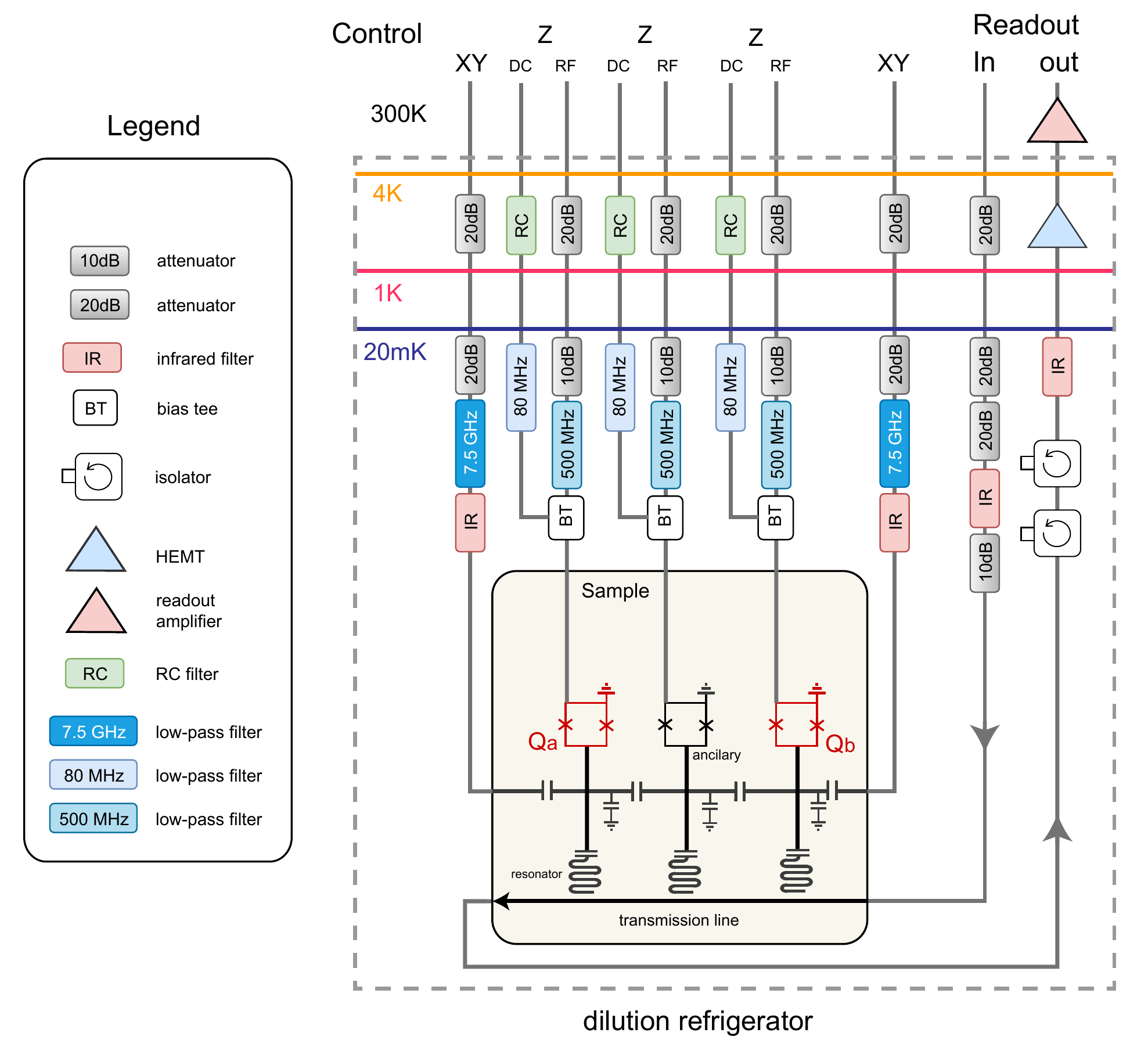}
		\caption{ Experimental setup. Frequency of a qubit is fine-tuned by a bias current on its Z control line. Microwave control pulses are applied to a qubit via its XY control line. Each qubit is capacitively coupled to a $\lambda/4$ resonator, which is coupled to a transmission line in turn. The state of a qubit can be deduced by measuring the transmission coefficient $S_{21}$ of the transmission line. The drawing at the bottom shows two qubits, $Q_a$ and $Q_b$, are effectively coupled via an ancillary qubit $Q_c$, which fine tunes the effective coupling strength (see the text for a detailed analysis).}\label{wiring}
	\end{figure*}

	Characteristic parameters of the superconducting quantum devices relevant to our experiment are summarized in Table I. Figure \ref{energy} shows the energy configuration of these devices. A schematic of the experimental setup, together with a drawing depicting the layout of the superconducting devices, are given in Fig.\ref{wiring}.	
	
	\begin{table}[!ht]
		\begin{tabular}{|cccccccc|} 
			\hline
			& $\omega_{ge}/2\pi$ $(\text{GHz})$ & $\alpha/2\pi~(\text{MHz})$ & $f_{r}$ $(\text{GHz})$ & $T_1^e(\mu s)$ & $T_1^f(\mu s)$ & $T_2^{ge}(\mu s)$ & $T_2^{ef}(\mu s)$  \\
			\hline \hline
			$Q_a$	& 5.520  & -278 & 6.828 & 10.1 & 9.4  & 1.8 & 1.9 \\
			\hline
			$Q_b$	& 5.633 & -270 & 6.884 & 8.7 & 6.7 & 0.8  &  0.8 \\
			\hline
		\end{tabular}
		\caption{Parameters of the qubits used in the experiment, including the frequency of the transition between the ground and first excited states $\omega_{ge}$, the anharmonicity $\alpha$ (see Fig.\ref{energy}(a) for its definition), frequency for readout $f_{r}$, and the relaxation and dephasing times $T_1^e$, $T_1^f$, $T_2^{ge}$, and $T_2^{ef}$.}
	\end{table}
	
	\clearpage
	
	\section{Tunable coupling of two qubits by an ancillary qubit}
	
\begin{figure*}
		\centering
		\includegraphics[width =0.8\textwidth]{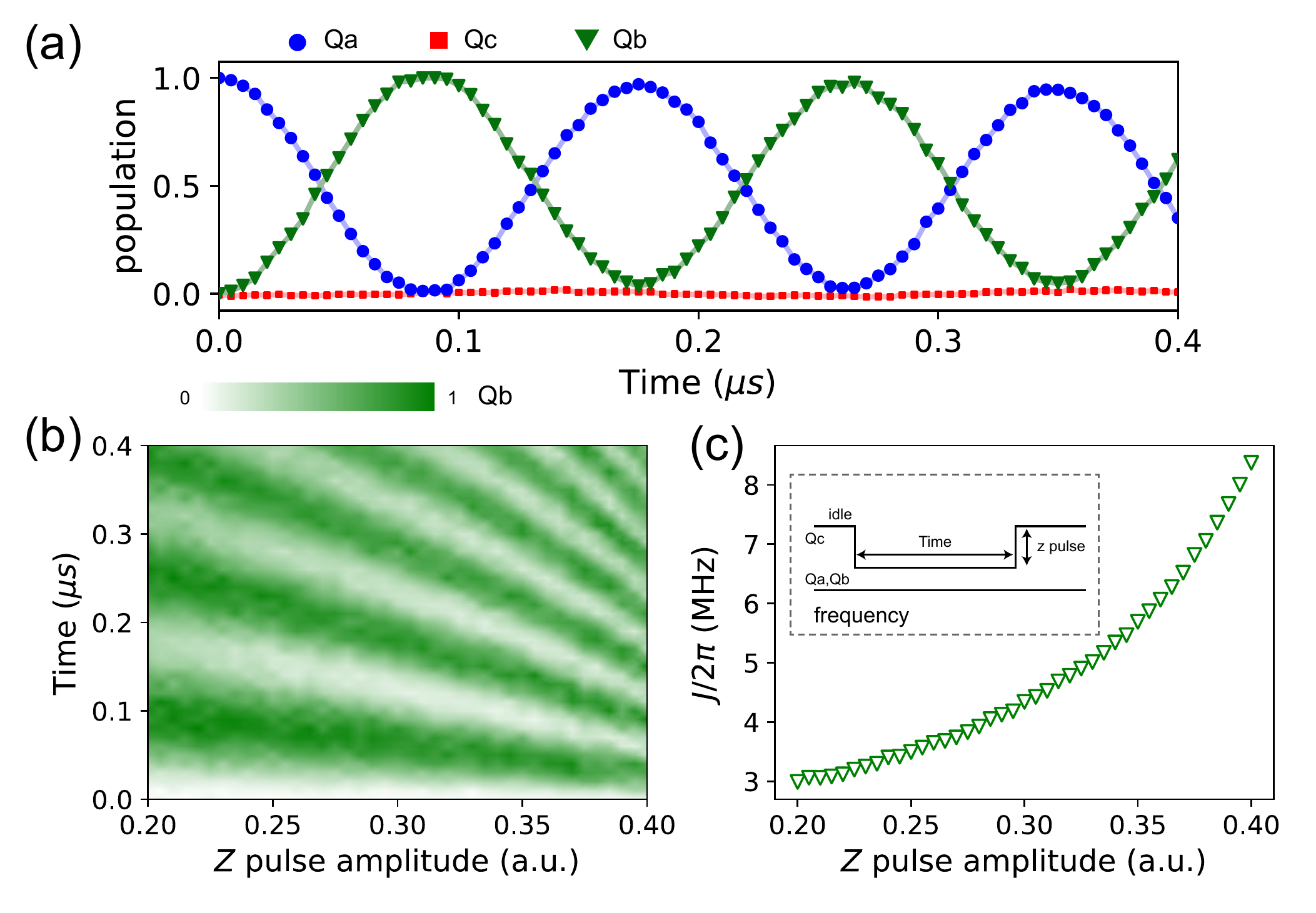}
		\caption{Effective coupling between two qubits using an ancillary qubit as a tunable coupler. (\textbf{a}) Measured populations of the excited state $P^e_{a,b,c}$ for $Q_{a,b,c}$, when the frequencies of the qubits $Q_a$ and $Q_b$ are tuned to be resonant $\bar{\omega}_{ge}^a = \bar{\omega}_{ge}^b$ and $Q_c$ is tuned away. The initial state is $\vert e \rangle_a \vert g \rangle_b \vert g \rangle_c$. (\textbf{b}) Measured population of $P^e_{b}$ as a fuction of time and Z pulse amplitude, where the Z pulse is applied to $Q_c$ to change $J$ by tuning the frequency detuning $\omega_{ge}^{a,b}-\omega_{ge}^c$. (\textbf{c}) The effective coupling $J$ extracted from (\textbf{b}), by fitting the data to a function of $P^e_{c}= A \cos (J t) + B$ with adjustable parameters $A$, $J$, and $B$. An increasing amplitude of the Z pulse decreases the frequency detuning $\omega_{ge}^{a,b}-\omega_{ge}^c$, thus increasing the effective coupling $J$.}\label{swapAB}
	\end{figure*}

	In the two-qubit experiment, resonant driving fields of the same frequency are applied to the two qubits $Q_a$ and $Q_b$. To suppress the effect of microwave crosstalk, $Q_a$ and $Q_b$ are effectively coupled through an ancillary qubit $Q_c$ with a strength of $J/2\pi$ around $6.7\text{ MHz}$ (see Fig.~\ref{energy}c and Ref.~\cite{Yan2018}). The Hamiltonian is given by
	\begin{equation}
	H = \sum_{j=a,b} \frac{1}{2}\omega^j_{ge}\sigma_z^j + \frac{1}{2}\omega_{ge}^c\sigma_z^c + \sum_{j=a,b} g_j(\sigma_+^j \sigma_-^c + \sigma_-^j \sigma_+^c)
	\end{equation}
	where $\sigma_z,\sigma_+,\sigma_-$ are the Pauli Z, raising and lowering operators defined in the eigenbasis of the corresponding qubit. The coupling strength is $g_{a,b}=25\textrm{ MHz}$. In a dispersive coupling regime where $g_j\ll \vert \Delta_{j}  \vert$ ($\Delta_{j} \equiv \omega_{ge}^j-\omega_{ge}^c$), we apply the Schrieffer-Wolff transformation $U=\exp(\sum_{a,b} [g_j/(\Delta_{j} )](\sigma_+^j \sigma_-^c - \sigma_-^j \sigma_+^c))$ and obtain an effective two-qubit Hamiltonian $H'=UHU^{-1}$ as
	\begin{equation}
	H = \sum_{j=a,b} \left(\frac{1}{2}\bar{\omega}^j_{ge}\sigma_z^j \right)+ \frac{g_a g_b}{\Delta}(\sigma_+^a \sigma_-^b + \sigma_-^b \sigma_+^a)
	\end{equation}
	where $\bar{\omega}^j_{ge}=\omega^j_{ge}+ g_j^2/\Delta_{j}$ and $1/\Delta=(1/\Delta_a + 1/\Delta_b)/2$.
	When the two qubits $Q_a$ and $Q_b$ are on resonance and the frequency of $Q_c$ is tuned away, the excitations of $Q_a$ and $Q_b$ can exchange $\vert e \rangle_a \vert g \rangle_b \leftrightarrow \vert g \rangle_a \vert e \rangle_b$ (Fig.~\ref{energy}c). Figure \ref{swapAB} shows the experimental data demonstrating an effective coupling between $Q_a$ and $Q_b$, using $Q_c$ as a tunable coupler.

	\clearpage

	\section{Closed-contour Hamiltonian in two-qubit subspace}
	Consider two superconducting qubits with a coupling strength of $J$, each subjected to a transverse resonant driving fields with an amplitude of $J/\sqrt{2}$. The phase difference between the two fields is $\phi_a-\phi_b = \phi$. The Hamiltonian of this system in the rotating frame of the frequencies of the qubits is
	\begin{equation}
	H_{2q}/\hbar= \frac{J}{\sqrt{2}}(\cos\phi\sigma^A_x + \sin\phi\sigma^A_y+\sigma^B_x) + \frac{J}{2}(\sigma_x\sigma_x+\sigma_y\sigma_y).
	\end{equation}
	After an evolving time of $t =2\pi/(3\sqrt{3}J) $, the two-qubit gates $U_{L},U_R$ for $\phi = \pi/2,-\pi/2$ are given by $e^{-iH t}$, and have the following forms:
	\begin{equation}
	U_L = \left(
	\begin{array}{cccc}
	\frac{1}{2} & 0 & -\frac{1}{\sqrt{2}} & \frac{i}{2} \\
	-\frac{i}{\sqrt{2}} & 0 & 0 & -\frac{1}{\sqrt{2}} \\
	0 & -i & 0 & 0 \\
	-\frac{i}{2} & 0 & -\frac{i}{\sqrt{2}} & \frac{1}{2} \\
	\end{array}
	\right),
	\end{equation}
	
	\begin{equation}
	U_R=\left(
	\begin{array}{cccc}
	\frac{1}{2} & -\frac{i}{\sqrt{2}} & 0 & -\frac{i}{2} \\
	0 & 0 & -i & 0 \\
	-\frac{1}{\sqrt{2}} & 0 & 0 & -\frac{i}{\sqrt{2}} \\
	\frac{i}{2} & -\frac{1}{\sqrt{2}} & 0 & \frac{1}{2} \\
	\end{array}
	\right).
	\end{equation}
	
	They can be decomposed into
	\begin{equation}
	U_L = U_1
	\left(
	\begin{array}{cccc}
	1 &  &  & \\
	&  & & -1 \\
	& -i &  &  \\
	&  & -i &  \\
	\end{array}
	\right)
	U_1^\dagger = U_1 U'_L  U_1^\dagger,
	\end{equation}
	
	\begin{equation}
	U_R = U_1^\dagger
	\left(
	\begin{array}{cccc}
	1 &  &  &  \\
	&  & -i &  \\
	&  &  & -i \\
	& -1 &  &  \\
	\end{array}
	\right)
	U_1 = U_1^\dagger U'_R  U_1,
	\end{equation}
	
	where $U_1$ is an iSWAP like gate
	\begin{equation}
	U_1 =
	\left(
	\begin{array}{cccc}
	\frac{1}{\sqrt{2}} &  &  & \frac{i}{\sqrt{2}} \\
	& 1 &  &  \\
	&  & 1 &  \\
	\frac{i}{\sqrt{2}} &  &  & \frac{1}{\sqrt{2}} \\
	\end{array}
	\right)
	\end{equation}
	We can transform the system into yet another frame by a unitary transformation $T = U_1^\dagger$, so $U_{L,R} \to U'_{L,R}$. Now the operations $U'_{L,R}$ acting on the four basis $\{0,1,2,3 \}$ are permutations of the three levels $1,2,3$: $U'_L:\{1,2,3\}\to\{2,3,1\}$ and $U'_R:\{1,2,3\}\to\{3,1,2\}$.
	
	Therefore, we find that the two-qubit system with driving fields is restricted in an invariant triplet subspace $\left \{\vert 10\rangle ,\vert 01\rangle , \left(\vert 00\rangle + e^{i\phi}\vert 11\rangle \right)/\sqrt{2}\right \}= \{\vert \bar{1}\rangle ,\vert \bar{2}\rangle ,\vert \bar{3}\rangle \} $, and has an effective Hamiltonian corresponding to a three-level system with a CCI:
	\begin{equation}
	\begin{split}
	H/\hbar&= J\left( \vert \bar{1} \rangle \langle \bar{2} \vert + \vert \bar{2} \rangle \langle \bar{3} \vert + e^{i\phi} \vert \bar{1} \rangle \langle \bar{3} \vert + H.c. \right)
	= J
	\begin{pmatrix}
	0&1&e^{i\phi}\\
	1&0&1\\
	e^{-i\phi}&1&0
	\end{pmatrix}
	\end{split}
	\end{equation}

	\section{Supplementary data}
	
	\begin{figure*}[h]
		\centering
		\includegraphics[width =1.0\textwidth]{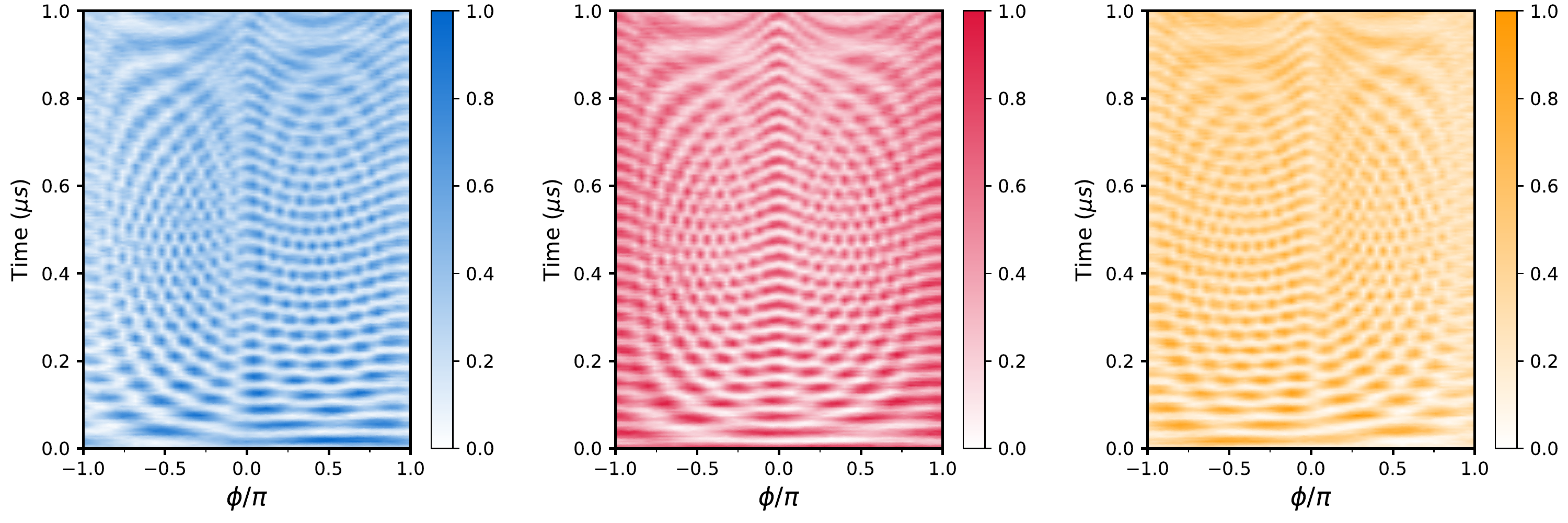}
		\caption{Experimental data used for the Fourier transform to obtain the energy spectrum of a single-qubit CCI (Fig.2 in the main text). For left to right: measured populations $P_{1}$, $P_{2}$, and $P_{3}$ as functions of the evolution time and $\phi$.}\label{supply_fig2b_Phi2D}
	\end{figure*}
	
	\begin{figure*}[h]
		\centering
		\includegraphics[width =0.6\textwidth]{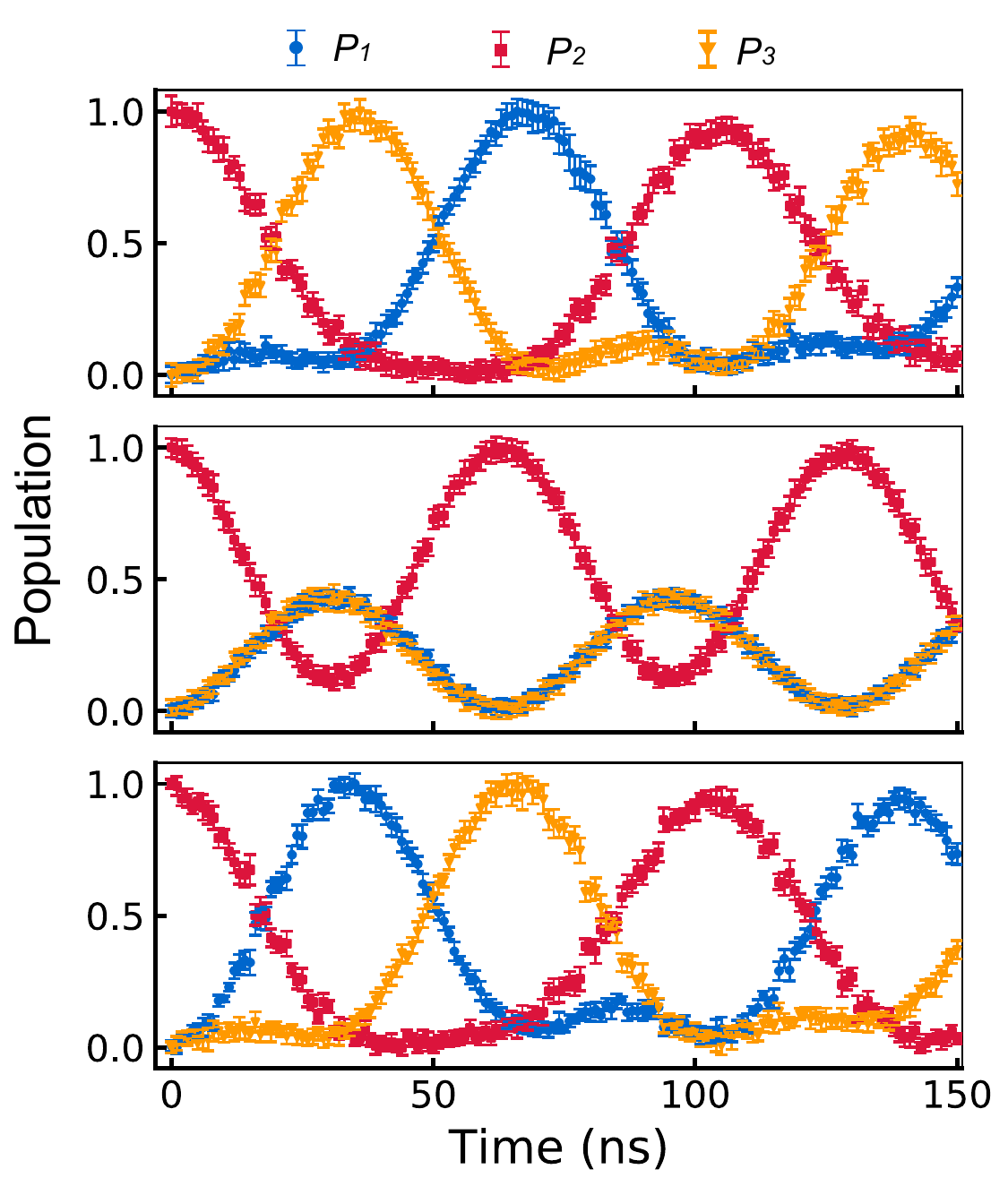}
		\caption{ Figure 2a in the main texts with errorbars. Each data point is obtained by 20 (upper and middel panels) or 10 (lower panel) repeated measurements, with each measurement containing 600 times of average. The errorbars represent the standard deviations of the corresponding 20 or 10 repeated measurements.}\label{supply_fig2a}
	\end{figure*}

	\begin{figure*}
		\centering
		\includegraphics[width =0.7\textwidth]{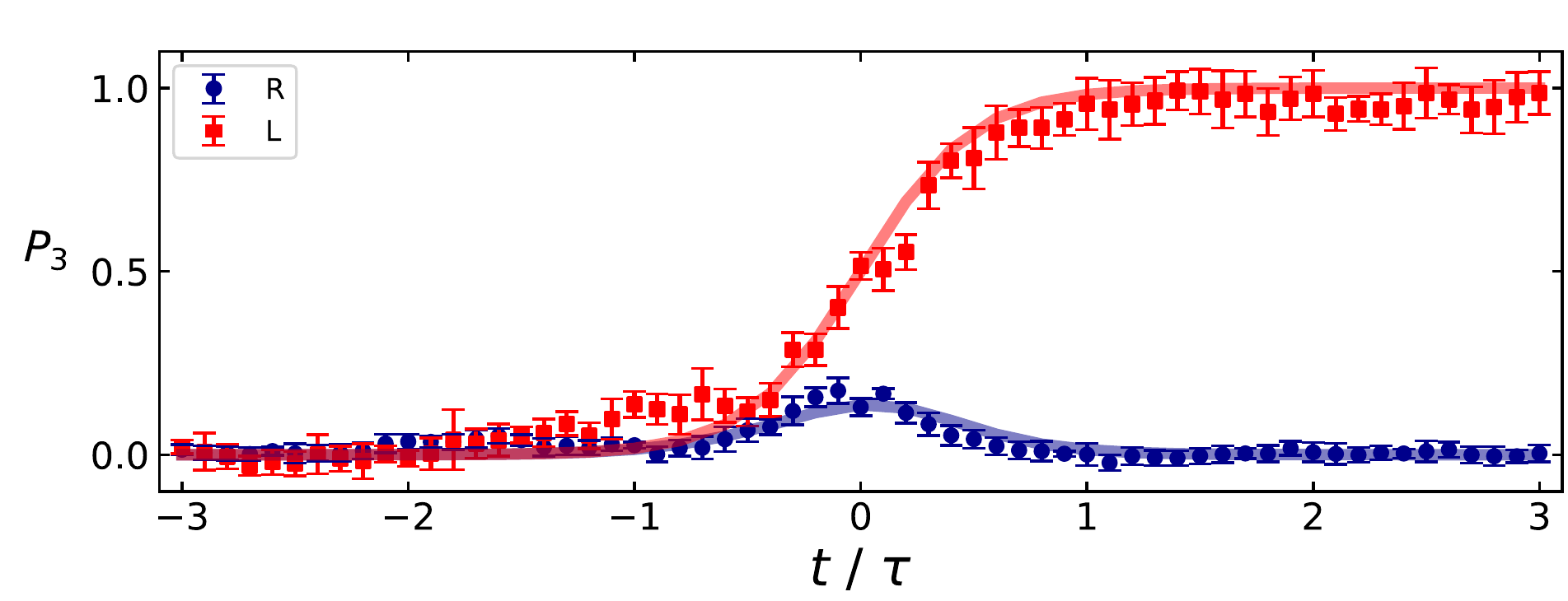}
		\caption{ Figure 3d in the main texts with errorbars. Each data point is obtained by 10 repeated measurements, with each measurement containing 600 times of average. The errorbars represent the standard deviations of the corresponding 10 repeated measurements.}\label{supply_fig3d}
	\end{figure*}
	
	\begin{figure*}
		\centering
		\includegraphics[width =0.6\textwidth]{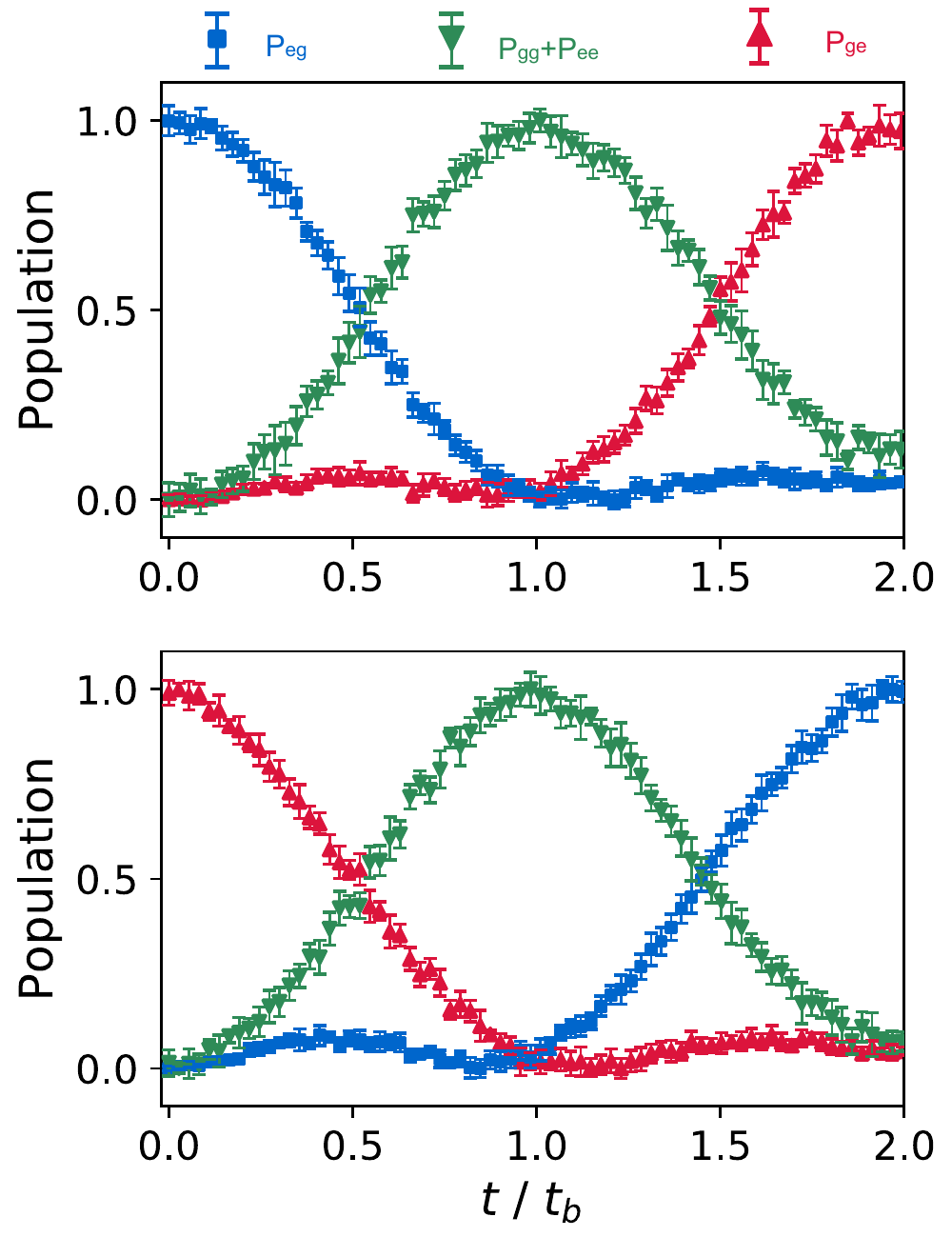}
		\caption{ Figure 4a and 4b in the main texts with errorbars. Each data point is obtained by 10 repeated measurements, with each measurement containing 600 times of average. The errorbars represent the standard deviations of the corresponding 10 repeated measurements.}\label{supply_fig4ab}
	\end{figure*}

	\clearpage
	
\end{document}